# A Fault Detection Scheme Utilizing Convolutional Neural Network for PV Solar Panels with High Accuracy

Mary Pa, Amin Kazemi
Department of Electrical Engineering, Lakehead University
Department of Mechanical and Industrial Engineering University of Toronto
Email: mpaparim@lakeheadu.ca, amin.kazemi@utoronto.ca

**Abstract**

Solar energy is one of the most dependable renewable energy technologies, as it is feasible almost everywhere globally. However, improving the efficiency of a solar PV system remains a significant challenge. To enhance the robustness of the solar system, this paper proposes a trained convolutional neural network (CNN) based fault detection scheme to divide the images of photovoltaic modules. For binary classification, the algorithm classifies the input images of PV cells into two categories (i.e. faulty or normal). To further assess the network's capability, the defective PV cells are organized into shadowy, cracked, or dusty cells, and the model is utilized for multiple classifications. The success rate for the proposed CNN model is 91.1% for binary classification and 88.6% for multi-classification. Thus, the proposed trained CNN model remarkably outperforms the CNN model presented in a previous study which used the same datasets. The proposed CNN-based fault detection model is straightforward, simple and effective and could be applied in the fault detection of solar panel.

## I- Introduction

Traditional fossil fuel-based power generations are shifting towards renewable energy to achieve net-zero greenhouse gas emissions by 2050. Renewable energy may be cheaper as well as be friendly to the environment. Examples of the most promising renewable energy sources are hydroelectric power, solar, and wind. Photovoltaic energy is one of the cleanest and most available renewable resources[1], [2], which has attracted much attention in recent decades [3]. Solar energy utilization is expected to increase more globally in the coming years. It is a promising alternative to fossil fuels and has a low adverse environmental impact. The use of solar energy can be downscaled to individual homes by using solar panels. These panels absorb the energy from the sun and provide power for a particular use, which makes the power system independent of larger electrical grids. Solar panels are usually designed to generate electricity in recent decades. However, they may face issues during their operation, which can reduce their efficiency or cause complete failure. Like any other electrical energy production system, photovoltaic power plants require monitoring and supervision to detect defects or abnormalities that may develop during operation and ensure their appropriate functioning and longevity while minimizing energy losses [4].The faults seen in a PV system can be grouped into several categories, such as a line-to-line defect[5] . The most common solar panel defects are the generation of a hot spot that causes degradation of the cells, microcracks due to thin construction, broken glass, and dust accumulation under the glass. All these defects may severely diminish the performance of the solar modules. The monitoring can be done on-site (*e.g.*[6]) or remotely (e.g.[7]). The author applied two CNN strategies to recognize issues in PV frameworks with a normal exactness of 73.5%, which isn't palatable and needs more improvement. A PV imperfection forecast approach was proposed by combining the fuzzy hypothesis and ANN and using voltage and power proportions as input factors to distinguish different PV issues. The author connected neural network methods for fault localization and classification of PV frameworks and reached good results even with noisy data [8].

Despite a lot of research in the intelligent algorithm-based fault detection of PV panels, determination of the best performing classifiers remains a challenge since their performances depend on various parameters such as the type of the problem, quality of the input signals or images, the number of inputs, number of layers, and the adjusting parameters in the networks. The current study provides a feature extraction and classification method based on a deep two- dimensional (2-D) CNN. An overview of the CNN-based fault detection algorithm is illustrated in Figure-1. Initially, the algorithm classifies the input images of PV cells into two simple categories, faulty or normal, called binary classification. Thereafter, the defective PV cells are further classified into shadowy, cracked, or dusty cells, known as multiple classifications. The approach used in this work is relatively simple while providing satisfactory outcomes. Moreover, the algorithm can be used to analyze several pictures of grid-connected solar PV panels and locate the faulty cells, which improves the durability and reliability of the PV systems.

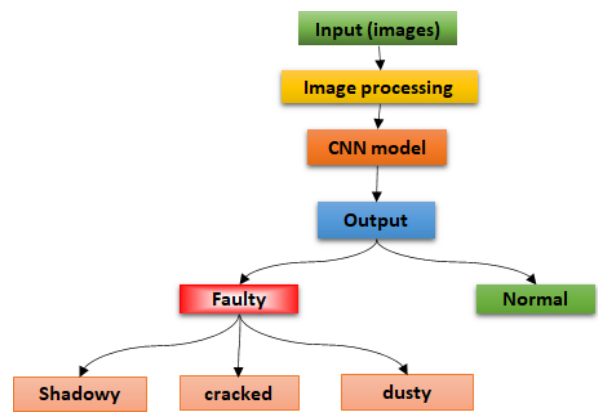

Figure-1 A general perspective of the CNN-based fault detection algorithm.

The convolutional neural network is a type of deep learning commonly utilized in image recognition and classification in remote sensing. The CNN transforms input information into numbers using several layers through its different topologies.

The general architecture of a CNN is demonstrated in Figure-2. The main parts of the CNN are generally convolution layers, pooling layers, fully connected layers, batch normalization, SoftMax, and classification layers which are briefly discussed below.

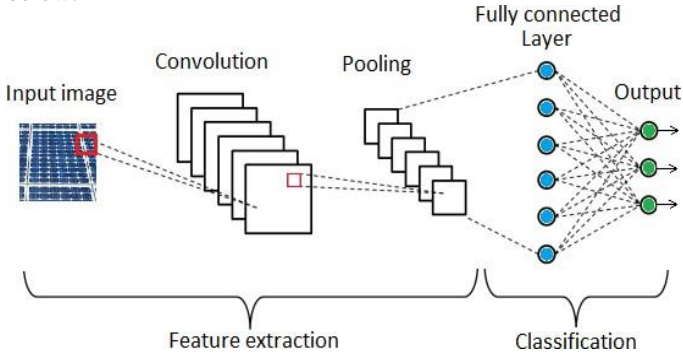

Figure-2 General architecture of the CNN.

*Convolution Layer* - The convolution layer extracts the image features and convolves the input image using convolution kernels of various sizes to produce the image's properties. It is followed by an aggregate process that combines the extracted features from diverse image parts [9].

$$s(t) = (x * w)(t) \quad \text{Eq-1}$$

$$s(t) = (x * w)(t) \sum x(a)w(t - a) \quad \text{Eq-2}$$

This equation indicates the complex function of t, *x* and *w*. The latter is the function of a two- dimensional network.

*Pooling layer* – This layer reduces the size of input images and makes the computation cost and duration much less. The layer connects the convolutional layer and the output of the model.

*Fully connected layer* - This layer maps the representation of inputs and outputs. It often has numeric values. The remaining of this chapter is organized as follows. Section II explains the proposed methodology. Section III provides the findings from the proposed CNN-based fault detection technique, and Section V contains the conclusion.

**II-Proposed CNN-based defect detection method**

The proposed CNN-based defect detection scheme is implemented using MATLAB software with the following system properties: CPU Intel ® Core™ i5-10400 CPU, 8 GB RAM with a 500 GB SSD hard disk, 64-bit operating system, and x64-based processor. This section explains the pre-treatment of the training dataset and the network's details for classification. The dataset consists of RBG images of solar panel arrays. The majority were collected from various internet search engines that offer photos of solar projects worldwide. Authors in [10]provided images of four classes (normal, cracked, dusty, and shadowed) of PV panels. The images are publicly available on www.github.com. The pictures of the solar cell modules are of the normal type, modules of cracked cells, images of dusty cells, and the rest are those partially covered by shadow. A typical illustration of each class is demonstrated in Figure -2. The data is separated into training (70%) and test sections (30%).

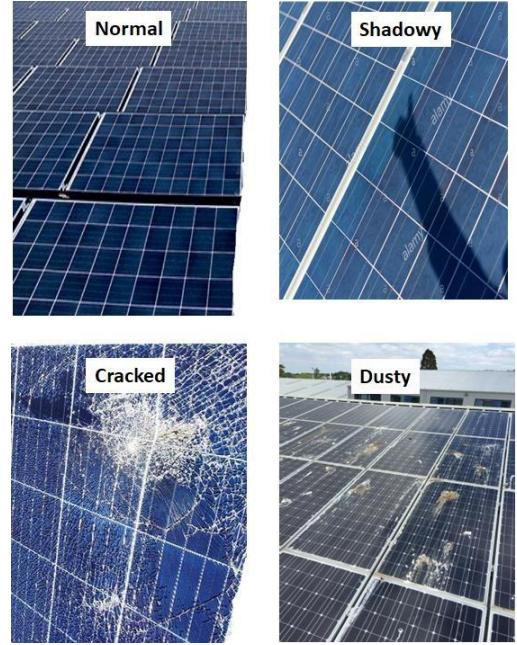

Figure-2 samples of images (cracked, dusty, shadowy and normal).

A segmentation model is utilized to extract an individual segmentation mask for each PV module, removing unnecessary information in the images and allowing for entirely accurate localization of PV modules in the pictures [11]. Then, data has been rescaled and normalized to be prepared for data augmentation to prevent overfitting. The adapted CNN classification framework and topology presented in this study can be seen in Figure-3 (a). In the proposed model, three layers of convolution followed by max-pooling to downscale the input data and extract the different features of the images at different levels. Besides, batch normalization layers are applied to robust the training procedure. In the fully connected layer, the extracted information from other neurons is combined and compared so that the network can predict the classes of each input image. The SoftMax layer is responsible for probability distribution over each possible class and classifying the datasets according to the most probable type. The network parameters and their assigned values are summarized in Figure-3 (a) and Figure-3 (b). Afterward, the proposed CNN is applied for training the data. For binary classification, the data sets are split into two different categories. 70% of the data are randomly selected for training purposes, whereas 30% of the remaining data was selected for testing and validation. The output of this classification network indicates the normal and faulty PV cells. Moreover, the parameters of the convolutional, pooling, and batch normalization layers are summarized.

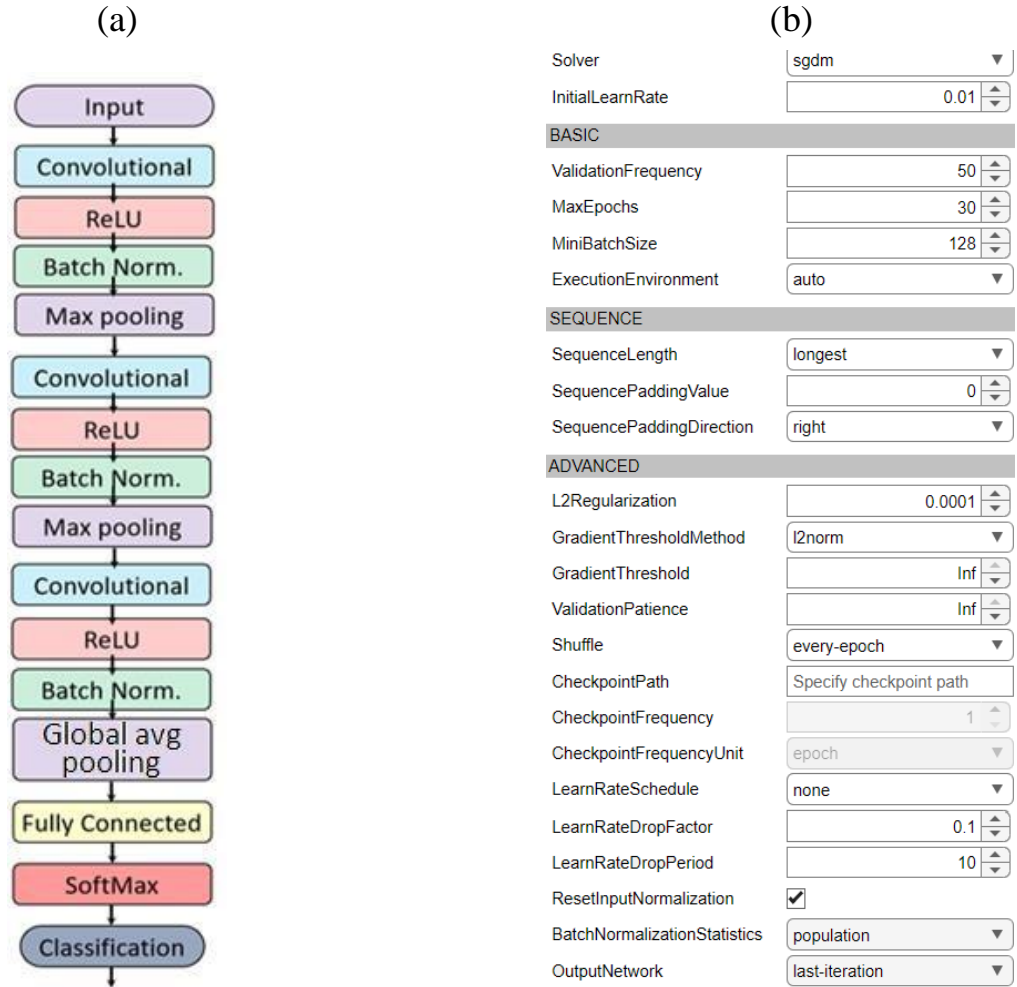

Figure -3 (a) CNN classification framework and topology, (b) network parameters and their assigned values

**III-Evaluation of the proposed CNN model**

The fault detection accuracy rate of the proposed CNN model is compared with another CNN model published in the literature [9] for binary and multi-classifications of the same dataset. According to the authors, for semantic segmentation, they used four convolutional layers, each of which was followed by a set of ReLu units and max pooling layers. After the last convolution layer, authors used a fully connected layer with SoftMax activation functions. The filter size they used was 3×3 pixels. For the multiclass architecture of the CNN, they used five convolutional layers which featured 5×5 filters. Each convolutional layer was followed by a batch normalization layer, a ReLu unit, and a max pooling layer[12].

The overall results of the comparison are shown in Table -1. It can be seen that the proposed CNN offers remarkably better accuracy (91.2%) than the other CNN model explored in ref [10], which was found to be 75.2%. The plots of accuracy and loss function with the epoch number for binary classification are demonstrated in Figure-4 The simulation was terminated after 30 epochs. However, it appears from the plots that even higher accuracies could be achieved if the simulation were allowed to run for more epochs.

Table-1 Comparison of the accuracies among various CNN models for binary classification

| Model | Overall accuracy |
|---|---|
| spinosa *et al*[9] | 75% |
| Proposed CNN model | 91 % |

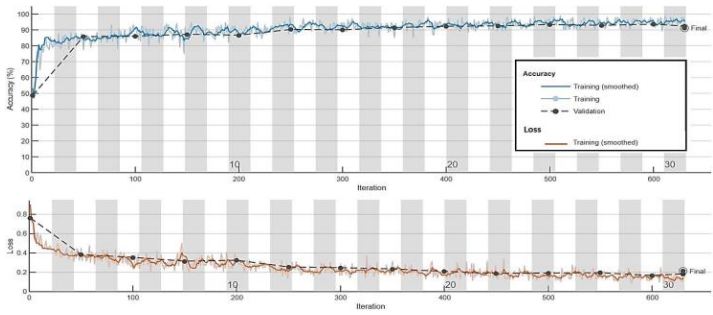

Figure-4 Comparative plots of accuracy and loss function with the epoch number for binary classification.

The CNN model was also evaluated by applying it to multiclass images. A comparison between the prediction accuracies of the CNN model with that proposed in ref. is summarized in Table -2. For the case of multi-classification, the model used in the current study outperforms the model in the literature by offering an accuracy of 88.6% compared to 70%. Similar to the binary case, the plots of accuracy and loss function shown in Figure-4 suggest that further improvements in the results could be achieved if the simulation ran for more epochs.

Table-2 Comparison of the accuracies among various CNN models for multi-classification

| Model | Overall accuracy |
|---|---|
| Espinosa et al[9] | 70 % |
| Proposed CNN model | 88.6 % |

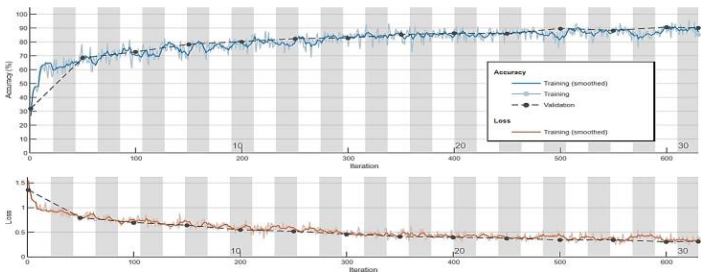

Figure-4 Comparative plots of accuracy and loss function with the epoch number for multi-classification.

The effectiveness of the proposed CNN model is again evaluated by reducing the number of layers. The model comprised of a convolutional removal step which is removed from the architecture previously shown in Figure-4(a), and the new architecture is demonstrated in Figure 5 (a). It can be seen in Figure-5 (b) and (c) that reducing the number of layers degrades the prediction accuracy. For the binary class, the model achieved an accuracy of nearly 80%, while for the multiclass, its accuracy is around 55%, both of which are less than the accuracies of the three-layer CNN model.

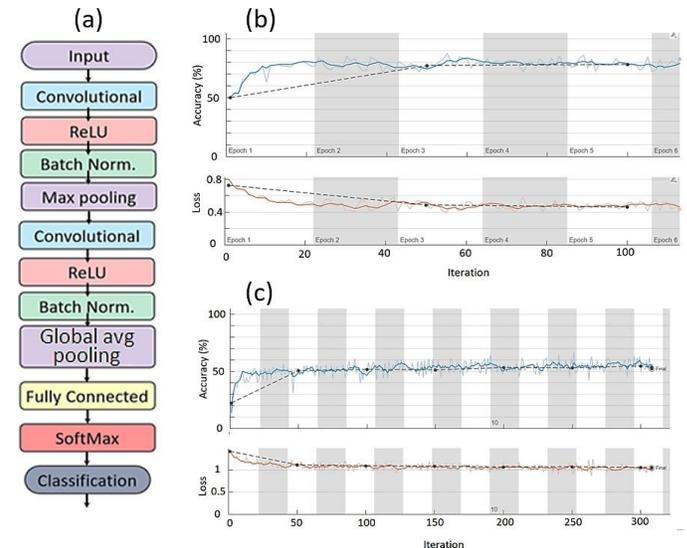

Figure-5 (a) CNN classification framework and topology with a reduced convolutional layer compared to the previous case. (b) Accuracy and loss function with epoch for binary class. (c) Accuracy and loss function with epoch for multiclass.

Transfer learning is a machine learning technique [13] that uses a pre-trained model or another task [3]. The capability of some frequently used pre-trained networks, including Squeeze Net, Dark Net, and Alex Net, in classifying the PV images is examined. For this The same dataset is used to re-train the pre-defined network available in the deep learning toolbox of MATLAB. A summary of the results is listed in Table -3.

Table-3 Accuracies of the prediction for three common CNN models.

| | Squeeze Net | Dark Net | Alex Net |
|---|---|---|---|
| Binary classification | 28% | 78% | 75% |
| Multi classification | 25% | 25% | 28% |

It can be observed that the pre-trained model did not have good performance as compared to the simple CNN model because the nature of the data is different than the datasets that these pre-trained models have been trained. Though, these pre-trained models may have much higher accuracies for current assessments with decreasing initial learning and changing optimizers and freezing their layers (using non-trainable layers).

**V-Conclusion**

A convolutional neural network-based defect detection of solar PV panel images has been presented in this thesis. The proposed technique may be utilized to improve the durability and reliability of the PV system and its operations. To rectify the challenges faced with the PV systems, this research study offers a deep learning-based solution through a simple fault detection system with a far less degree of complexity than other alternatives, so the overall efficiency of the PV system is not affected. The network was used to classify the images of PV panels divided into four classes of normal, cracked, dusty, and shadowed. It was found that the proposed CNN technique effectively classifies data sets with an accuracy of 91.2% for two classes (normal and faulty) and 88.6% for four categories (normal, cracked, dusty and shadowed), respectively, based on the experimental data. The performance of the proposed model was also compared with a model in the literature that used the same dataset. It was found that our CNN model outperformed the model in the literature by 16% for the binary and 18.6% for multi-class. Transfer learning was used to investigate if some pre-trained models can perform better. The results suggested that those models are not suitable for this specific dataset and did not provide acceptable accuracies. The presented model can be easily applied to other similar engineering applications, such as the inspection of wind turbines in which a manual check is not safe. Furthermore, the proposed algorithm is very flexible for implementation. Thus, the proposed fault detection scheme could be applied to a real- life solar farm to increase the reliability of the solar system and decrease the maintenance cost of PV panel.

## References

[1] S. R. Madeti and S. N. Singh, "Monitoring system for photovoltaic plants: A review," *Renewable and Sustainable Energy Reviews*, vol. 67, pp. 1180–1207, Jan. 2017, doi: 10.1016/J.RSER.2016.09.088.

[2] M. Rachid and M. M. Khafallah, "HAJAR DOUBABI," Reims, Jul. 2021.

[3] A. Triki-Lahiani, A. Bennani-Ben Abdelghani, and I. Slama-Belkhodja, "Fault detection and monitoring systems for photovoltaic installations: A review," *Renewable and Sustainable Energy Reviews*, vol. 82, pp. 2680–2692, Feb. 2018, doi: 10.1016/J.RSER.2017.09.101.

[4] R. Dabou *et al.*, "Monitoring and performance analysis of grid connected photovoltaic under different climatic conditions in south Algeria," *Energy Convers Manag*, vol. 130, pp. 200–206, Dec. 2016, doi: 10.1016/J.ENCONMAN.2016.10.058.

[5] C. Kopacz, S. Spataru, D. Sera, and T. Kerekes, "Remote and centralized monitoring of PV power plants," in *2014 International Conference on Optimization of Electrical and Electronic Equipment, OPTIM 2014*, 2014, pp. 721–728. doi: 10.1109/OPTIM.2014.6851005.

[6] C. Ventura and G. M. Tina, "Utility scale photovoltaic plant indices and models for on-line monitoring and fault detection purposes," *Electric Power Systems Research*, vol. 136, pp. 43–56, Jul. 2016, doi: 10.1016/j.epsr.2016.02.006.

[7] S. T. Kebir, N. Cheggaga, A. Ilinca, and S. Boulouma, "An Efficient Neural Network-Based Method for Diagnosing Faults of PV Array," 2021, doi: 10.3390/su13116194.

[8] R. K. Mandal, N. Anand, N. Sahu, and P. Kale, "PV system fault classification using SVM accelerated by dimension reduction using PCA," *PIICON 2020 - 9th IEEE Power India International Conference*, Feb. 2020, doi: 10.1109/PIICON49524.2020.9112896.

[9] A. Rico Espinosa, M. Bressan, and L. F. Giraldo, "Failure signature classification in solar photovoltaic plants using RGB images and convolutional neural networks," *Renew Energy*, vol. 162, pp. 249–256, Dec. 2020, doi: 10.1016/J.RENENE.2020.07.154.

[10] F. Aziz, A. Ul Haq, S. Ahmad, Y. Mahmoud, M. Jalal, and U. Ali, "A Novel Convolutional Neural Network-Based Approach for Fault Classification in Photovoltaic Arrays," *IEEE Access*, vol. 8, pp. 41889–41904, 2020, doi: 10.1109/ACCESS.2020.2977116.

[11] A. Ghanbarpour, A. H. Mahmoud, and M. A. Lill, "Instantaneous generation of protein hydration properties from static structures," *Commun Chem*, vol. 3, no. 1, pp. 1–19, Dec. 2020, doi: 10.1038/s42004-020-00435-5.

[12] O. Day and T. M. Khoshgoftaar, "A survey on heterogeneous transfer learning," *J Big Data*, vol. 4, no. 1, Dec. 2017, doi: 10.1186/s40537-017-0089-0.

[13] K. Weiss, T. M. Khoshgoftaar, and D. Wang, "A survey of transfer learning," *J Big Data*, vol. 3, no. 1, p. 9, Dec. 2016, doi: 10.1186/s40537-016-0043-6.